
\magnification=\magstep1
\centerline{\bf Event horizons and apparent
horizons in spherically symmetric geometries.}
\vskip 2cm

\centerline{Edward Malec}
\centerline{Physics Department, UCC, Cork, Ireland}
\vskip 0.5cm
\centerline{and}
\vskip 0.5cm
\centerline{Institute of Physics, UJ}
\centerline{300-59  Cracow, Reymonta 4, Poland}
\vskip 2cm
\centerline{PACS numbers}
\centerline{\bf Abstract.}
\vskip 0.5cm

 Spherical configurations that are very massive must
be surrounded by apparent  horizons. These in turn,
when placed outside a collapsing body,
must propagate outward with a velocity equal to the
velocity of radially outgoing photons. That proves,
within the framework of (1+3) formalism
and without resorting to the Birkhoff theorem, that
apparent horizons coincide with event horizons.

\vfill \eject

There exists  a gap in our understanding of
spherically symmetric geometries. From the Birkhoff
theorem$^1$ we know that once the areal radius
{\bf R} of a
charged collapsing body becomes equal
to $m+\sqrt{m^2-q^2}$, where $m$ is the
asymptotic
(Einstein - Freund - Arnowitt - Deser - Misner) mass
and $q$ is the total charge, then the body hides
within an event horizon that coincides with
a sphere of the areal radius {\bf R}. At the same time
there is no direct proof that in the  alternative
dynamical
(1+3) description of a spherical collapse
an event horizon
develops.
\par The intention of this letter is to fill this gap.
Let the spherically symmetric metric line element be

$$ds^2=-\alpha^2(r, t)dt^2 +a(r, t)dr^2
+b(r, t)r^2d\Omega^2.\eqno(1)$$

Let the boundary of a collapsing body be a sphere of
a coordinate radius $r_0$. Let us define  $c=
{{-q^2+m^2}\over 4}$.
First, let us notice that there exists
a static solution outside a collapsing body$^2$

$$a=b=(1+{m\over r}+{c\over { r^2}})^2,$$

$$\mid \alpha  \mid =
 \mid {{r-\sqrt{c}}\over {r+\sqrt{c}}}
\mid^{{m\over {2\sqrt{c}}}}\mid r^2-c\mid^{{{q^2}
\over {2(r^2+mr+c}}}e^{{{q^2}\over 2}\int_r^{\infty }
{{\ln (\mid s^2-c^2\mid )(2s+m)}\over {(s^2+ms+c)^2}}ds}
.$$

{}From this one readily infers that at a coordinate
radius $r=\sqrt{c}$ (that is, at an areal radius
${\bf R}=m+\sqrt{m^2-q^2}$) exists an event
 horizon. In the case of  vanishing total charge $q$
the above solution coincides with the Schwarzschild
solution in isotropic coordinates$^1$.
That is a trivial situation that I will not discuss
in what follows.
\par
Let us assume that a  spherically symmetric spacetime
generated by a collapsing body exists (i. e., the global
Cauchy evolution exists) and that it
can be foliated by maximal slices that are asymptotically
flat. It was proven elsewhere$^3$ that when the amount
of matter minus a total radial momentum exceeds a multiple
( 1 or 7/6)  of the proper radius, then apparent
horizons must form.
\par Let us  assume that there exists an
apparent horizon outside a
(neutral or charged) collapsing body of
a compact support. (We do not
exclude  longe-ranged potentials, with an
energy density  like the electrostatic Coulomb-like
energy in vacuum, that might exist outside a body.
That is, we allow for an electrovacuum,
but we definitely exclude everything else.)
$t$ is now a parameter that labels maximal slices; it
coincides with a proper time of an
external observer that is localized very far
from a collapsing body. $r$ is a coordinate radius
and $r^2 d\Omega^2 $ is a standard 2-sphere
metric element. The maximality condition
implies that components of the extrinsic curvature
satisfy the following equations

$$K^r_r={{\partial_t a}\over {2a\mid \alpha \mid }}=
-2K^{\phi}_{\phi }= -2K^{\theta }_{\theta }=
 -{{\partial_tb}\over b\mid \alpha \mid }. \eqno(2)$$

Under these conditions, one proves that the
Penrose$^{4,5}$ inequality (which actually becomes an
equality) holds true, that at the surface of an apparent
horizon

$$m = \sqrt{S\over {16\pi }} +
q^2\sqrt{{\pi }\over S}.\eqno(3)$$

\par
It will be convenient to prove (3)
in an isotropic system of coordinates in which
$a=b=\phi^4$. It is easy to prove that this form of a
metric can be achieved just by performing
a suitable change of a radial coordinate on a fixed
Cauchy
slice. Morever, the final result - equation (3) - is
already expressed in a coordinate independent way.
 The proof goes as follows.
In electro-vacuum the hamiltonian  constraint reads$^3$

$$\hat \Delta \phi = -\hat E_i\hat E^i \phi^{-3}
- {{ K_{ij} K^{ij}\phi^5} \over 8}.\eqno(4)$$

Here the hatted quantities refer to the flat background
metric and $\hat E^r = {q\over {r^2}}, E^{\theta }=
E^{\phi }=0.$ From the momentum constraints one gets$^3$

$$K_{ij}= (n_in_j-{{g_{ij}}\over 3}){C\over
{\phi^6R^3}}, \eqno(5)$$

where $n_{j}$ is a normal vector in the physical
non-hatted metric.
 C is  a constant, a multiple of
the total radial momentum of the collapsing system.
\par
Equation (4) has a conserved ($r-$independent) quantity

$$E={r\over 8}(2r\partial_r\phi +\phi )^2-
{{r\phi^2}\over 8} - {{q^2}\over {8r^2\phi^2}}-
{{C^2}\over {72 r^2 \phi^6}}. \eqno(6)$$

Assuming asymptotic flatness one finds that $E=-{m\over
4}$, where $m$ is the asymptotic
 mass.
Notice also that $r\phi^2$ is equal to an areal
radius, $r\phi^2=\sqrt{{S\over {4\pi }}}$.
 After some rearrangements one might
write (6) as follows

$$m - [\sqrt{S\over {16\pi }} +
q^2\sqrt{{\pi }\over S}] =-{r\over 2}[2r\partial_r \phi
+\phi  -{C\over {3r^2\phi^3}}][2r\partial_r \phi
+\phi  +{C\over {3r^2\phi^3}}].\eqno(7)$$

If $S$ is an apparent horizon, then the first bracket
on the right hand side of (7) vanishes, which proves (3).
\par
Equation (3) actually holds on all
maximal slices as far as the apparent horizons
remains outside the collapsing body. That means
that the areal radius

$${\bf R}=r\sqrt{b}\eqno(8)$$

(we are coming back to the original
metric notation (1)) of the apparent horizon
must be conserved in time, since the mass $m$ is
conserved in time in asymptotically flat systems.
That is, the full time derivative of {\bf R} must vanish,
which leads to the following equality

$$V\sqrt{b}{{2b+r\partial_r b}\over {2b}}+
r{{\partial_tb}\over {2\sqrt{b}}}=0.\eqno(9)$$

Here $V=dr/dt$ is the coordinate velocity expansion of
the apparent horizon.
Now, the condition for the apparent horizon reads

$$2b+r\partial_r b={{brK_{rr}}\over {\sqrt{a}}};\eqno(10)$$

inserting this into (9) and using (2) we obtain

$${{\sqrt{b} K_{rr}r}\over {1\sqrt{a}}}
[V- {{\mid \alpha \mid }\over\sqrt{a}}]=0.\eqno(11)$$

We conclude that
the apparent horizon expands with a radial velocity

$$V={{\mid \alpha \mid }\over {\sqrt{a}}}.\eqno(12)$$

But, from  equation (1) we know that
${{\mid \alpha \mid}\over {\sqrt{a}}}$ is equal to the
velocity of radially
outgoing photons. Therefore, no material
object can escape from within the apparent horizon;
 it just coincides with  the event
horizon.
 We can say more.
Actually, the equation (7)
might be interpreted in the following
way: if an areal radius of a sphere $S$ satisfies (7),
then $S$ is an apparent horizon. By continuity,
all Cauchy slices must contain surfaces satisfying (7),
and they must remain in vacuum, since they move
with the velocity of light.
That means that, as long as maximal slices exist and
under conditions stated previously, the apparent horizon
must exist forever and it coincides with an event horizon.
In this way we have proven
in the framework of (1+3) formalism, without resorting
to the Birkhoff theorem, that there
exists an event horizon that intersects
each Cauchy slice along a sphere of the areal
radius ${\bf R}= m+\sqrt{m^2-q^2}$.
\par
 The above conclusion
 combined with one of theorems of $^3$, allows
one to formulate the following statement:
\par
{\bf Theorem.} If
 on an initial Cauchy hypersurface with
momentarily static initial data

$$M>L,\eqno(13)$$

where $M$ is a total mass$^1$ and $L$ is a proper radius
of a collapsing body,
then
the whole space-time history of the body must contain an
event horizon that surrounds the body.
\par Thus, if the energy content inside a ball of a
fixed radius becomes large, then it hides under an event
horizon. That proves a version of the Cosmic Censorship
Hypothesis$^6$ (CCH) in which $singularities$ are supplied
with a qualifier $massive$; {\bf massive singularities
are hidden under an event horizon}- this is a version
of CCH that looks plausible.
\par
The above results are easily proved in Schwarzschild
coordinates, thanks to the possibility of
applying the Birkhoff theorem.
 But, the standard way  of solving that
particular problem in Schwarzschild
coordinates  does not
suggest us any possible way of dealing with nonspherical
geometries. (As a parenthetical remark, let us
point  that one needs also to prove that actually
Schwarzschild coordinates are diffeomorphic with
those used above, which of course cannot be done
if apparent horizons are present.)
 \par In contrast with that, the present investigation
 bases on the (1+3) splitting of spacetime and
 its real value  relies on the fact that it reveals
a  way to treat nonspherical geometries.
In particular, it demonstrates  that
the concept of {\it trapped surfaces} introduced
by Penrose
is valid; that the {\it Penrose inequality}
might be useful; and that, finally,
the existence of {\it asymptotically flat maximal slices}
might be coherently related with CCH.
\par
The application of the above ideas to more general
classess of
collapsing systems will be reported elsewhere$^7$.
\par Acknowledgements.
I wish to thank Niall O'Murchadha for
many discussions and valuable remarks, that greatly
contributed to my understanding of problems discussed here.
This work was supported in part by the Polish Government
Grant KBN 2526/92.
\vfill \eject
\centerline{References.}
\vskip 1cm
\par
$^1$ e. g., Misner, Wheeler, Thorne, {\it Gravitation}
Freeman 1973;\par
$^2$ I am convinced that this form
of the electrovacuum solution of Einstein
-Maxwell equations has been obtained
hitherto, but I was not able to find any reference;\par
$^3$ P. Bizo\'n, E. Malec, N. O' Murchadha,
{\it Phys. Rev. Lett.} {\bf 61}, 1147(1988);\par
$^4$ R. Penrose, p. 631 in {\it Seminar on Differential
Geometry}, Princeton University Press 1982;\par
$^5$ G. Gibbons, p. 194 in {\it Global Riemannian
Geometry}, ed. N. J. Willmore and N. J. Hitchin,
Elli Horwood, New York 1984;\par
$^6$ R. Penrose, {\it Riv. Nuovo Cimento} {\bf 1}, 252
(1969);\par
$^7$ E. Malec, N. O' Murchadha, work in progress.
 \vfill \eject

 \end